\documentclass[aps,twocolumn]{revtex4}
\usepackage{graphpap}
\usepackage[dvips]{graphicx}
\usepackage[dvips]{graphics}
\usepackage{color}

\begin{document}

\title{Transport through a strongly coupled graphene quantum dot in perpendicular magnetic field}
\author{J. G\"uttinger}
\email{guettinj@phys.ethz.ch}
\author{C. Stampfer}
\altaffiliation{Present address: JARA-FIT and II. Institute of Physics, RWTH Aachen, 52074 Aachen, Germany}\affiliation{Solid State Physics Laboratory, ETH Zurich, 8093 Zurich, Switzerland}
\author{T. Frey}
\affiliation{Solid State Physics Laboratory, ETH Zurich, 8093 Zurich, Switzerland}
\author{T. Ihn}
\affiliation{Solid State Physics Laboratory, ETH Zurich, 8093 Zurich, Switzerland}
\author{K. Ensslin}
\affiliation{Solid State Physics Laboratory, ETH Zurich, 8093 Zurich, Switzerland}

\date{ \today}

\begin{abstract}
We present transport measurements on a strongly coupled graphene quantum dot in a perpendicular magnetic field. The device consists of an etched single-layer graphene flake with two narrow constrictions separating a 140 nm diameter island from source and drain graphene contacts. Lateral graphene gates are used to electrostatically tune the device. Measurements of Coulomb resonances, including constriction resonances and Coulomb diamonds prove the functionality of the graphene quantum dot with a charging energy of $\approx 4.5~$meV. We show the evolution of Coulomb resonances as a function of perpendicular magnetic field, which provides indications of the formation of the graphene specific $0^{th}$ Landau level. Finally, we demonstrate that the complex pattern superimposing the quantum dot energy spectra is due
to the formation of additional localized states with increasing magnetic field.

 \end{abstract}

 \maketitle

\newpage
\section{Introduction}

Graphene~\cite{nov04,gei07}, a two-dimensional solid consisting of carbon atoms arranged in a honeycomb lattice has 
a number of
unique electronic properties~\cite{cas09} such as the gapless linear dispersion, and
the unique Landau level (LL) spectrum~\cite{nov05,zha05}. The low atomic weight of carbon and the low nuclear spin concentration, arising from the $\approx 99$\% natural abundance of $^{12}$C, promises weak spin orbit and hyperfine coupling. This makes graphene a promising material for spintronic devices~\cite{Awsch07,Tomb07} and spin-qubit based quantum computation~\cite{Loss98,Elzer04,Petta05,Trau07}. Additionaly, the
strong suppression of electron backscattering ~\cite{nov05,zha05} makes it interesting for future high mobility nanoelectronic applications in general~\cite{kat07, Avouris07}. 
Advances in fabricating
graphene nanostructures have helped
to overcome intrinsic difficulties in (i) creating tunneling
barriers and (ii) confining electrons in bulk graphene, where
transport is dominated by Klein tunneling-related phenomena~\cite{dom99, kat06}. Along this route graphene nanoribbons~\cite{han07,che07,dai08,sta09,mol09,tod09,liu09} and quantum dots~\cite{sta08a,pon08,sta08aa,sch09,gue09,gue09a,gue08,ihn10} have been fabricated. Coulomb blockade~\cite{sta08a,pon08,sta08aa}, quantum confinement effects~\cite{sch09,gue09,gue09a} and charge detection~\cite{gue08} have been reported.
Moreover, graphene
nanostructures may allow to investigate phenomena related
to massless Dirac Fermions in confined dimensions~\cite{pon08,ber87,sch08,rec09,lib10,lib09,you09}. In general, the investigation of signatures of graphene-specific properties in
quantum dots is of interest to understand the
addition spectra, the spin states and dynamics of confined
graphene quasi-particles.

Here, we report on tunneling spectroscopy (i.e. transport)
measurements on a 140~nm graphene quantum dot with open barriers, which can be tuned by
a number of lateral graphene gates~\cite{mol07}. In contrast to the measurements reported in Ref.~\onlinecite{gue09} the more open dot in the present investigation enables us to observe Coulomb peaks with higher conductance and the larger dot size reduces the magnetic field required to see graphene specific signatures in the spectra.
We characterize the graphene quantum dot device focusing on
the quantum dot Coulomb resonances which can be distinguished from additional resonances 
present in the graphene tunneling barriers.
We discuss the evolution of a number of Coulomb
resonances in the vicinity of the charge neutrality point in a perpendicular magnetic
field from the low-field regime to the regime where Landau
levels are expected to form. In particular, we investigate the
device characteristics at elevated perpendicular magnetic fields, where
we observe the formation of multiple-dots giving rise to
(highly reproducible) complex patterns in the addition spectra.

 \begin{figure*}\centering
\includegraphics[draft=false,keepaspectratio=true,clip,%
                   width=1\linewidth]%
                   {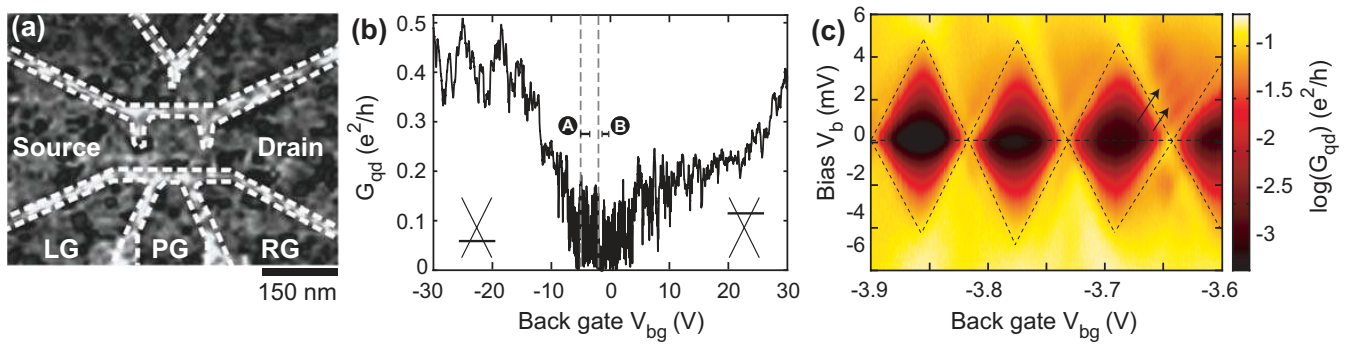}                  
\caption[FIG1]{(color online)
(a) Scanning force microscopy of the graphene quantum dot device. The overall chemical potential of the device is tuned by a global back gate, where as the right side gate (rg) is used for local asymmetric tuning. The extension of the dot is around 140~nm with 75~nm wide and 25~nm long constrictions. The white dashed lines delineating the quantum dot perimeter are added for clarity. (b) Measurement of the source (S)-drain (D) conductance for varying back gate voltage showing a transport gap from around -5 to 3~V ($V_\mathrm{b} = 200~\mu$V). (c) Coulomb diamond measurements in the gap showing a charging energy of around 4.5~meV. This energy is lower than what has been measured in an other dot of similar size (Ref.~\cite{sch09}), most likely because of the increased coupling to the leads. The arrows point to faint lines outside the diamonds. The extracted energy difference of around 1~meV is a reasonable addition energy for excited states. Note that for the measurement in (c), in addition to the BG the right side gate was changed according to $V_{\mathrm{rg}} = -0.57\cdot V_{\mathrm{bg}}-1.59$~V.} 
\label{trdansport}
\end{figure*}

\section{Device fabrication}

The fabrication process of the presented graphene nano-device is based on the mechanical exfoliation of (natural) graphite by adhesive
tapes [24, 25, 28]. The substrate material consists of highly
doped silicon (Si$^{++}$) bulk material covered with 295~nm of silicon oxide (SiO$_2$), where thickness
(and roughness) of the SiO$_2$ top layer is crucial for the Raman~\cite{dav07a} and scanning force microscope based identification
of single-layer graphene
flakes. Standard photolithography followed by metalization
and lift-off is used to pattern
arrays of reference alignment markers on the substrate
which are later used to re-identify the locations of individual
graphene flakes on the chip and to align further processing
patterns.
The graphene flakes are structured to submicron
dimensions by electron beam lithography (EBL) and reactive ion etching based techniques in order to 
fulfill the nanodevice design requirement. 

After etching and removing the residual resist, the
graphene nanostructures are contacted by an additional
EBL step, followed by metalization and lift-off. 

A scanning force microscope image of the final device studied
here is shown in Fig.~1a. The approximately 140~nm diameter graphene quantum dot is connected to
source (S) and drain (D) via two graphene constrictions
with a width of $\approx 75$~nm and a length of $\approx 25$~nm, both acting as tunneling barriers.
The dot and the leads can be further tuned by the highly
doped silicon substrate used as a back gate (BG) and three
in-plane graphene gates: the left side gate (LG), the plunger gate (PG) and 
the right side gate (RG). Apart from the geometry, the main difference of this sample compared to the device presented in Ref.~\onlinecite{gue09} is the higher root mean square variation of the height ($r_{\mathrm{h}}$) on the island. While there are no visible resist residues on the island of the sample in Ref.~\onlinecite{gue09} with $r_{\mathrm{h}} \approx 0.35~$nm, there are many dot-like residues on the sample presented here giving $r_{\mathrm{h}} \approx 1.1~$nm.

\section{Measurements}

All measurements have been performed at a base temperature of T = 1.8~K 
in a variable temperature cryostat. We
have measured the two-terminal conductance through the
graphene quantum dot device by applying a symmetric
DC bias voltage $V_\mathrm{b}$ while measuring the source-drain current through
the quantum dot with a noise level below 10~fA.
For differential conductance measurements a small AC
bias, $V_\mathrm{b,ac} = $100~$\mu$V has been superimposed on $V_\mathrm{b}$ and the
differential conductance has been measured with lock-in
techniques at a frequency of 76~Hz.

In Fig. 1b we show the conductance $G_\mathrm{qd}$ as a
function of back gate voltage at low bias ($V_\mathrm{b}$ = 200 $\mu$V)
highlighting the strong suppression of the conductance
around the charge neutrality point (-5 $< V_\mathrm{bg} <$ 3 V)
due to the so-called transport gap~\cite{tod09, sta09, mol09, liu09}. Here we tune
transport from the hole to the electron regime,
as illustrated by the left and right inset in Fig.~1b. 
The large number of resonances with amplitudes in the range of up to
$0.1$~$e^2/h$
inside the gap region may be due to both, (i) resonances
in the graphene constrictions acting as tunneling barriers
[4] (and thus being mainly responsible for the large
extension of this transport gap) and (ii) Coulomb resonances
of the quantum dot itself (see also examples of Coulomb diamonds in Fig.~1c). 
At room temperature these resonances disappear and 
a conductance value of $0.76$~$e^2/h$ is measured at $V_\mathrm{bg}$ = 0~V. 

\begin{figure}\centering
\includegraphics[draft=false,keepaspectratio=true,clip,%
                   width=1\linewidth]%
                   {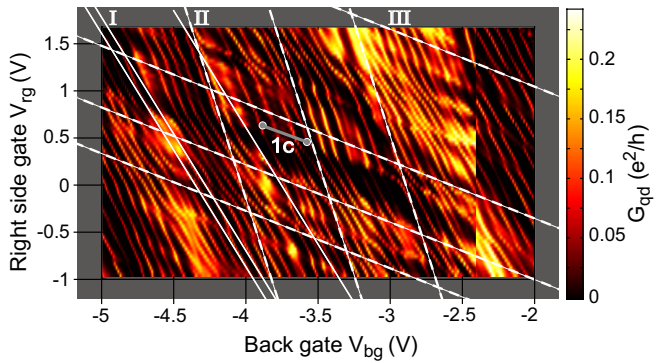}                  
\caption[FIG2]{(color online)
Conductance of the quantum dot with varying right gate and back gate voltage measured at bias voltage $V_\mathrm{b}=200~\mu$V. Coulomb resonances and modulations of their amplitude with different slopes are observed (dashed white lines). The extracted relative side gate back gate lever arms are $\alpha_\mathrm{rg/bg}^{(I)} \approx 0.4$, $\alpha_\mathrm{rg/bg}^{(II)} \approx 0.2$ and $\alpha_\mathrm{rg/bg}^{(III)} \approx 1.65$. Lever arm (III) is attributed to resonances in the right constriction which are strongly tuned by the right side gate. In contrast resonances with lever arm (II) are only weakly affected by the right side gate and therefore attributed to states in the left constriction. The periodic resonances marked with (I) are attributed to resonances in the dot in agreement with the intermediate slope.} 
\label{fig2}
\end{figure}

\subsection{Coulomb blockade measurements at B = 0 T}

By focusing on a smaller back gate voltage range within the
transport gap (indicated by the dashed lines in Fig.~1b) 
and measuring the conductance as a function of $V_\mathrm{bg}$ and the
right side gate $V_\mathrm{rg}$ much more fine-structure appears, as shown 
in Fig.~2.
A large number of resonances is observed with sequences of diagonal lines
(see white lines in Fig.~2) with different slopes, corresponding to different
lever arms ($\alpha$'s).
By sweeping the right side gate ($V_\mathrm{rg}$) we break the left-right symmetry of the transport
response (see also Fig.~1a). This allows us to distinguish between
resonances located either near the quantum dot or the left and right constriction.
The steeper the slope in Fig.~2 the less this resonance can be electrostatically tuned by the right
side gate and, consequently, the larger the distance between the corresponding localized state and the right side gate.
Subsequently, the steepest slope (II, corresponding to $\alpha^{(II)}_\mathrm{rg/bg}$~=~0.2) can be attributed to resonances
in the left constriction and the least steepest slope (III, $\alpha^{(III)}_\mathrm{rg/bg}$~=~1.6) belongs to resonances
in the right constriction. Both are highlighted as white dashed lines in Fig.~2.
The Coulomb resonances of the quantum dot appear with an intermediate slope (I, $\alpha^{(I)}_\mathrm{rg/bg}$~=~0.4) and exhibit clearly
the smallest spacing in back gate voltage, $\Delta V_\mathrm{bg} \approx$~0.1~V. This is a good indication that they belong to
the largest charged island in the system, which obviously is the 140~nm large graphene quantum dot,
which is much larger than the localized states inside the graphene constrictions acting as tunneling barriers.

Corresponding Coulomb diamond measurements~\cite{kou97}, that is, measurements of the differential conductance
as a function of bias voltage $V_\mathrm{b}$ and $V_\mathrm{bg}$ (i.e. $V_\mathrm{rg}=-0.57\cdot V_\mathrm{bg}-1.59$~V)
have been performed along the (diagonal) solid gray line in Fig.~2 and are shown in Fig.~1c. From the extent of these diamonds in bias direction we estimate the average charging energy of the graphene quantum dot to be $E_\mathrm{c} = 4.5$~meV, which is in reasonable agreement with the size of the graphene quantum dot~\cite{sta08a,sta08aa,sch09}. Moreover, we observe faint strongly broadened lines outside the diamonds running
parallel to their edges, as indicated by arrows in Fig.~1c. The extracted energy difference of roughly 1~meV is reasonable for electronic excited states in this system~\cite{sch09}.

\begin{figure*}\centering
\includegraphics[draft=false,keepaspectratio=true,clip,%
                   width=0.7\linewidth]%
                   {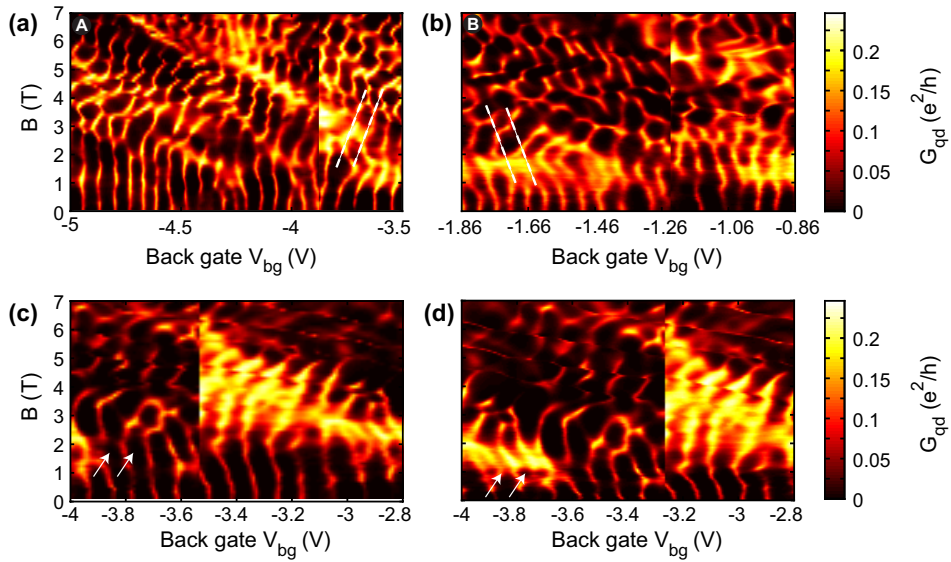}                  
\caption[FIG3]{(color online)
Evolution of Coulomb peaks under the influence of a magnetic field in different gate voltage regimes ($V_{bias} = 200~\mu$V). (a) More on the hole side. (b) more on the electron side. In contrast to (a) $V_\mathrm{rg} = -2.15$~V is applied to the right gate in (b). The effect of the right gate to the dot is taken into account in the back gate scale to allow comparison with Fig.~1b. (c,d) Reproducibility of the measurement for different magnetic field sweep directions (0 - 7~T in (c), 7 - 0~T in (d)). The right side gate is changed according to $V_{\mathrm{rg}} = -0.57\cdot V_{\mathrm{bg}}-1.59$~V (see Fig.~2), with an applied bias of $V_\mathrm{b} = 200~\mu$V.  }
\label{fig3} 
\end{figure*}

\subsection{Coulomb resonances as a function of a perpendicular magnetic field}
In Fig.~3 we show a large number of Coulomb resonances as function of a magnetic
field perpendicular to the graphene sample plane.
The measurement shown in Fig.~3a has been taken in the back gate voltage range from
$V_\mathrm{bg}=$-5 to -3.5~V, at $V_\mathrm{rg}=0~$V (highlighted by the horizontal line (A) in
Fig.~1b).
Thus we are in a regime where transport is dominated by holes (i.e. we are at the left
hand side of the charge neutrality point in Fig.~1b), which is also confirmed
by the evolution of the Coulomb resonances in the perpendicular magnetic field as shown in Fig.~3a.
There is a common trend of the resonances to bend towards higher energies (higher $V_\mathrm{bg}$)
for increasing magnetic field, in good agreement with Refs.~\cite{gue09,gue09a,sch08,rec09,lib10}.
The finite magnetic field introduces an additional length scale $\ell_\mathrm{B} = \sqrt{\hbar/e B} \approx 25~$nm$/\sqrt{B [T]}$ which competes with the diameter $d$ of the dot. Therefore the ratio $d/\ell_\mathrm{B}$ is a relevant parameter for the observation of Landau levels in graphene quantum dot devices.
Here, the comparatively large size ($d \approx$~140~nm) of the dot promises an increased spectroscopy window for studying the onset and the formation of Landau levels in graphene quantum dots in contrast to earlier work~\cite{gue09,gue09a} (where $d \approx$~50~nm).
Moreover, we expect that in larger graphene quantum dots, where the surface-to-boundary ratio increases edge effects should be less relevant.
In Figs.~3a,c,d we indeed observe some characteristics of the Fock-Darwin-like spectrum~\cite{sch08,rec09,lib10}
of hole states in a graphene quantum dot in the near vicinity of the charge neutrality point: (i) the levels stay more or less at constant energy (gate voltage) up to a certain B-field, where (ii) the levels feature a kink, whose B-field onset increases for increasing number of particles, and (iii) we observe that the levels convergence 
towards higher energies (see white dashed lines in Fig.~3a). The pronounced kink feature (see arrows in Figs.~3c,d) indicate filling factor $\nu$~=~2 in the 
quantum dot.
However, this overall pattern is heavily disturbed by additional resonances caused by localized states, regions of 
multi-dot behavior, strong amplitude modulations due to constriction resonances and a large number of additional crossings, which are not
yet fully understood.
This becomes even worse when investigating the electron regime
(see horizontal line (B) in Fig.~1b), as shown in Fig.~3b.
Individual Coulomb resonances can (only) be identified for low magnetic fields B $<$ 2~T and a slight tendency for their bending towards lower energies
might be identified (please see white dashed lines in Fig.~3b). 
For magnetic fields larger than 3~T it becomes very hard to identify individual Coulomb resonances in the complex and reproducible
conductance pattern.

In order to demonstrate the reproducibility of these complex patterns we show an up (Fig.~3c) and a
down (Fig.~3d) sweep of the very same $B$-$V_\mathrm{bg}$ parameter space.
These two measurements, have different resolution and thus different sweep rates in both
the $B$ and $V_\mathrm{bg}$ direction. However, all the individual features are highly reproducible (but hard to
understand) despite the fact that we find some small hysteresis in magnetic field for $B < 3$~T (see white arrows in Figs.~3c,d).
The origin of the complex patterns shown in Fig.~3 can be understood when having a closer look at 
charge stability diagrams (such as Fig.~2) for different magnetic fields.

In Fig.~4a we show an example of a sequence of dot Coulomb resonances in the $V_\mathrm{rg}$-$V_\mathrm{bg}$ plane.
The slope corresponding to $\alpha^{(I)}_\mathrm{rg/bg}~\approx~0.4$ and the spacing of $\Delta V_\mathrm{bg} \approx 0.1$~V are in good agreement with Fig.~2 and lead to the conclusion that we observe single quantum dot behavior over a large parameter range.
However, if we measure the current in the very same $V_\mathrm{rg}$-$V_\mathrm{bg}$ parameter space at B~=~7~T the
pattern changes significantly and the diagonal lines are substituted by a strong hexagonal pattern (see dashed lines) typical for two coupled quantum dots~\cite{wiel02}. The two states forming the hexagone pattern show relative lever arms of $\alpha^{(I)}_\mathrm{rg/bg}~\approx~0.4$ and $\alpha^{(IV)}_\mathrm{rg/bg}~\approx~1$. While the resonances with $\alpha^{(I)}_\mathrm{rg/bg}$ are attributed to the original dot, $\alpha^{(IV)}_\mathrm{rg/bg}$ corresponds to a new and strongly coupled localization formed close to the right constriction. Additional resonances from the right constriction with $\alpha^{(III)}_\mathrm{rg/bg}~\approx~1.65$ (see above) are still visible.

We interpret the magnetic field dependence in the following way. At low but increasing magnetic field we see in almost all measurements an increase of the conductance through the dot (see e.g. Fig.~3). Assuming diffusive boundary scattering such a conductance onset in magnetic field occurs due to reduced backscattering~\cite{bvh91} and 
has been observed in other measurements on graphene nanoribbons~\cite{oost10, bai10}. The maximum conductance is reached around $B~\approx~1.5$~T corresponding to a magnetic length $\ell_B = \sqrt{\hbar/eB} \approx 50~$nm in rough agreement with the size of the constriction. As the magnetic field is further increased the complex pattern with many crossings starts to emerge, attributed to the formation of additional quantum dots around the right constriction with strong coupling to the original dot. The formation of such localized puddles is understood as a consequence of the increased magnetic confinement where $\ell_B$ is getting smaller than the extension of potential valleys induced by disorder.

\begin{figure}\centering
\includegraphics[draft=false,keepaspectratio=true,clip,%
                   width=1\linewidth]%
                   {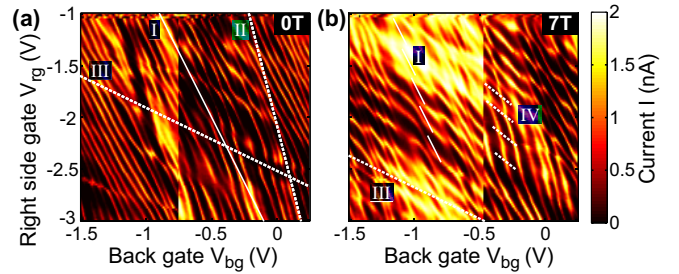}                  
\caption[FIG4]{(color online)
Dot conductance as a function of right gate and back gate voltage at a magnetic field of (a) 0~T and (b) 7~T. The spectrum is dominated by dot resonances marked with the solid line in (a) with a relative lever arm of $\alpha_\mathrm{rg/bg}^{(I)} \approx 0.4$ (see also Fig.~2). (b) At a magnetic field of 7~T a hexagon pattern with two characteristic slopes is observed. Their corresponding lever arms are $\alpha_\mathrm{rg/bg}^{(I)} \approx 0.4$ attributed to the dot and $\alpha_\mathrm{rg/bg}^{(IV)} \approx 1$ origin around the right constriction. 
} 
\label{fig4}
\end{figure}

\section{Conclusion}
In summary we have presented detailed studies of transport through an open and larger graphene quantum dot (compared to Ref.~\onlinecite{gue09}) in the vicinity of the charge neutrality point as a function of perpendicular magnetic field. The evolution of Coulomb resonances in a magnetic field showed the signatures of Landau level formation in the quantum dot. Indications for the crossing of filling factor $\nu$ = 2 are obtained by the observation of "kinks" in spectral lines before bending towards the charge neutrality point. However the observation is disturbed by the formation of a pronounced additional localized state at high magnetic fields in the vicinity of the right constriction. 
Although the use of open constrictions enhances the visibility of the Coulomb peaks and reduces the transport-gap region, emerging pronounced parasitic localized states make the analysis very difficult. 
For a further in-depth analysis of the addition spectra around the electron-hole crossover it is hence beneficial to minimize the amount of disorder and to use clearly defined constrictions. These should be thin compared to the dot diameter in order to get different energy scales for quantum dot resonances and constriction resonances, which are easy to distinguish. However, the constrictions need to be wide enough to enable conductance measurements around the electron-hole crossover without a charge detector.

{Acknowledgment ---}
The authors wish to thank F. Libisch, 
P. Studerus, C. Barengo, F. Molitor and S. Schnez
for help and discussions. Support by the ETH FIRST
Lab, the Swiss National Science Foundation and NCCR
nanoscience are gratefully acknowledged.



\end{document}